\documentclass{PoS}

\title{Advanced search for the extension of unresolved TeV sources
  with H.E.S.S. \\{\normalsize First measurement of the extension of
    the Crab nebula at TeV energies}} 

\newcommand{\hess}{H.E.S.S.}

\ShortTitle{Advanced search for the extension of unresolved TeV sources with H.E.S.S.}

\author{\speaker{M. Holler}$^{a}$, D. Berge$^{b}$, J. Hahn$^{c}$, D. Khangulyan$^{d}$, R.~D. Parsons$^{c}$, for the H.E.S.S. collaboration\\
        ${}^{a}$ Institut f\"ur Astro- und Teilchenphysik, Universit\"at Innsbruck\\
        ${}^{b}$ GRAPPA, Anton Pannekoek Institute for Astronomy, University of Amsterdam\\
        ${}^{c}$ Max-Planck-Institut f\"ur Kernphysik\\
        ${}^{d}$ Department of Physics, Rikkyo University\\
        E-mail: \email{markus.holler@uibk.ac.at}}

      \abstract{The resolution power of current Imaging Atmospheric
        Cherenkov Telescopes is presently restricted to scales of a
        few arcminutes. In the very high-energy (VHE; $E > 100$\,GeV)
        gamma-ray regime, the measurement of source sizes that are
        comparable to or smaller than the resolution of the instrument
        is usually limited by statistics and in particular by the
        uncertainties in the characterisation of the instrument Point
        Spread Function (PSF). The PSF varies strongly with
        observation and instrument conditions demanding time dependent
        simulations of these conditions. Employing such simulations,
        we substantially improve our understanding of the
        H.E.S.S. PSF and are now able to probe source extensions well
        below one arcminute scale. We present the results of this new
        approach applied to known VHE gamma-ray sources and show how
        this enables us to reveal for the first time the extension of
        the Crab nebula at TeV gamma-ray energies, with a width of
        $52''$ assuming a Gaussian source shape.}

\FullConference{35th International Cosmic Ray Conference - ICRC2017\\
		12-20 July, 2017\\
		Bexco, Busan, Korea}

\begin{document}

\section{Instrument and Dataset}
\hess\ is an array of five Imaging Atmospheric Cherenkov Telescopes
(IACTs) located in the Khomas Highlands of Namibia. The Crab nebula
has been observed with the \hess\ telescopes since the beginning of
operations in 2004. For the measurement presented here we only use
observations with the four IACTs of the first phase of \hess\ In
addition to the standard run-quality selection
criteria~\cite{2006_HessCrab}, we apply additional cuts e.g.\ on the
maximum wind speed ($< 3\,\mathrm{m}/\mathrm{s}$), the observation
wobble offset ($< 0.8^{\circ}$), and the zenith angle of the
observations ($< 55^{\circ}$) to define a high-quality dataset for
extension measurements. The resulting Crab nebula dataset consists of
observations performed between February 2004 and November 2011,
amounting to a total deadtime-corrected live time of $25.7\,$h. Since
the Crab nebula is a Northern source, the typical observation zenith
angles at the \hess\ site in the Southern hemisphere are
$45^{\circ}-50^{\circ}$.

\section{Analysis and Results}
The data analysis is performed using semi-analytical air-shower
templates~\cite{2009_deNaurois}.  To further improve the angular
resolution, we use a tight cut on the direction reconstruction
uncertainty for the event selection. Furthermore, we demand energies
of $E > 0.7\,$TeV to eliminate potential systematic effects near the
energy threshold of the dataset\footnote{The \hess\ energy threshold
  for the Crab nebula is relatively large because the source is only
  visible under large zenith angles.}. We detect the Crab nebula with
a statistical significance of $137\sigma$ (using Eq.~$17$ from
\cite{1983_LiMa}) and obtain a signal to background ratio of $58$
within $0.1^{\circ}$ around the source position. We find a total
number of about $4600$ excess events, the majority of them with
energies between the lower cut of $0.7\,$ and $10\,$TeV.
\begin{figure*}
  \centering
  \includegraphics[width=\textwidth]{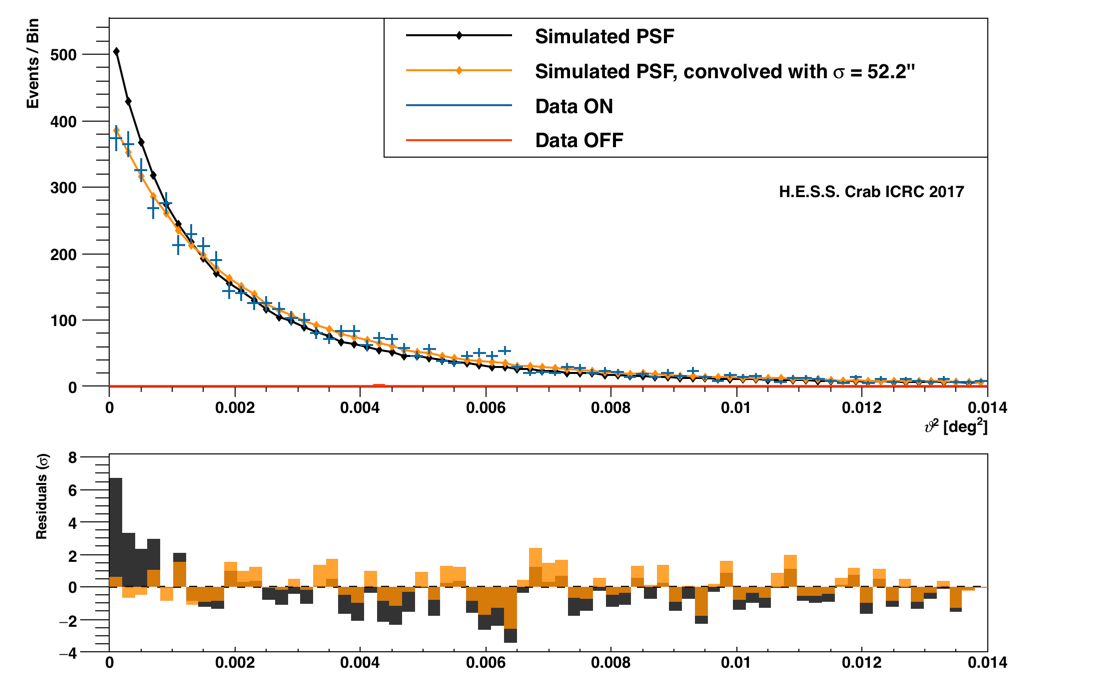}
  \caption{\textit{Top:} $\vartheta^2$ histogram of gamma-ray-like
    events from the Crab nebula ($ON$) along with background events
    from displaced sky regions ($OFF$, scaled to the $ON$ region). The background events are barely visible in the plot due to the large $S/B$ ratio. For comparison, the simulated
    PSF and the PSF convolved with the best-fit Gaussian are shown as
    well. \textit{Bottom:} Significance of the bin-wise deviation
    ($MC-data$) of the measured excess when compared to the plain PSF
    (black) and the convolved one (orange).}
  \label{fig_theta2}
\end{figure*}
We show the resulting distribution of events as a histogram in squared
angular distance ($\vartheta^2$) in the \textit{top} panel of
Fig.~\ref{fig_theta2}. The value of $\vartheta^2$ is calculcated with
respect to the centroid of the gamma-ray excess count distribution in
the sky. The centroid position in equatorial coordinates (J2000) is
$\alpha = 5\mathrm{h}34\mathrm{m}30.9\mathrm{s} \pm
(1.2\mathrm{s})_{\mathrm{stat}} \pm
(20\mathrm{s})_{\mathrm{sys}}$,
$\delta = +22^\circ00'44.5'' \pm 1.1''_\mathrm{stat} \pm
20''_\mathrm{sys}$ (systematic error from \cite{2004_Gillessen}), at a distance of $16.8''$ from, and within
uncertainties compatible with, the Crab pulsar location.

Dedicated run-wise Monte-Carlo (MC) simulations of the dataset,
including the actual instrument and observation conditions at the time
the data were recorded, are generated as described
in~\cite{2017_RWS_ICRC}. We re-weight the simulations to mimic the
energy spectrum of the Crab nebula and analyse them with the same
algorithms and analysis configurations as the actual data. The
resulting $\vartheta^2$ histogram of this MC analysis serves as the
PSF for this source and dataset and is shown in the upper panel of
Fig.~\ref{fig_theta2}. The $68\%$, $80\%$, and $90\%$ containment
radii of our PSF are $0.05^\circ$, $0.07^\circ$, and $0.09^\circ$,
respectively. 

As shown in Fig.~\ref{fig_theta2}, the PSF is highly inconsistent with the
distribution of the gamma-ray excess counts. The residuals in the
lower panel of Fig.~\ref{fig_theta2} indicate a clear broadening of the
data compared to the simulated PSF. To study this further, we perform
a 2D morphology fit with \textit{Sherpa}~\cite{2001_Sherpa}, using the
sky images of gamma-ray-like events around the Crab nebula, of
gamma-ray-like background events estimated from a ring well outside
the source~\cite{2007_Berge_Background}, as well as the simulated
PSF. To determine the best fit, we convolve the PSF with a 2D Gaussian
with different widths. For each width, we calculate a likelihood value
to assess the compatibility of the data and the convolved PSF. We find
the best-fit extension to be
$\sigma_{2\mathrm{D,G}} = 52.2'' \pm 2.9''_{\mathrm{stat}} \pm
7.8''_{\mathrm{sys}}$, with a preference of an extension of the Crab
nebula over a point-source assumption of
$\mathrm{TS} \approx 83\,$\footnote{$\mathrm{TS}$ is the likelihood
  ratio test statistic and $\sqrt{\mathrm{TS}}$ can be interpreted as
  statistical significance.}. As systematic uncertainty of the
extension we quote the quadratic sum of uncertainties related to the
calibration and analysis method, to the spectral shape used to re-weight
the MC PSF, and to the fit method.

The resulting best-fit convolution is also plotted in
Fig.~\ref{fig_theta2}. It clearly provides a good description of the
data both in the upper panel and the residuals in the lower panel.

To demonstrate the robustness of our results, we apply the same
analysis using time-dependent simulations to two other bright and
highly significant extragalactic gamma-ray sources, the active
galactic nuclei PKS$~$2155-304 and Markarian$~$421.
\begin{figure*}
  \centering
  \includegraphics[width=0.8\textwidth]{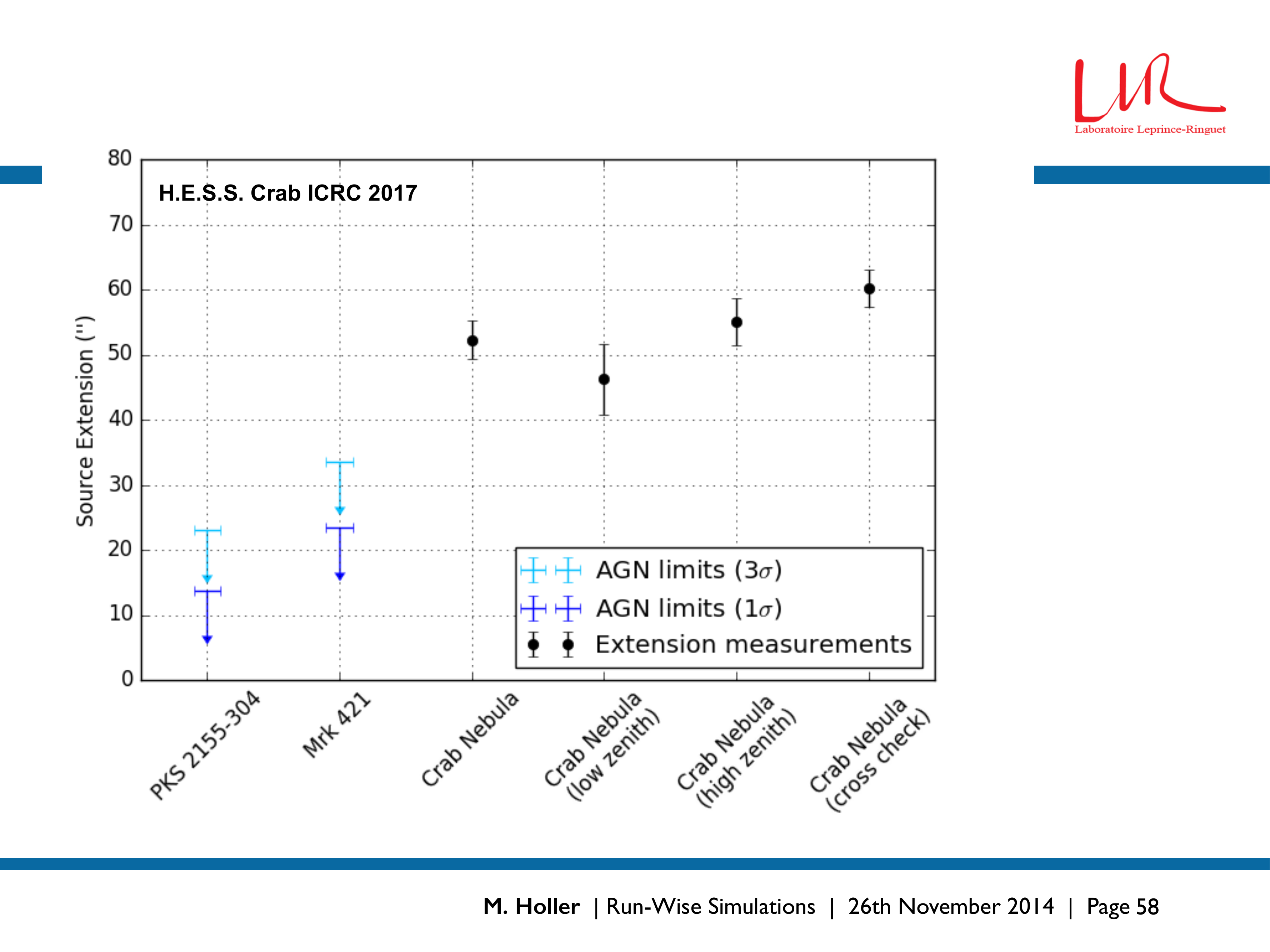}
  \caption{Overview of the derived extension upper limits of PKS
    2155-304 and Markarian 421, as well as the measured extension of
    the Crab nebula and systematic checks. The low and high zenith angle band correspond to $44-46^{\circ}$ and $46-55^{\circ}$, respectively.}
  \label{fig_uls_ext}
\end{figure*}
As illustrated in Fig.~\ref{fig_uls_ext}, we find both sources
compatible with being point-like and show upper limits on their
extension. In both cases, these limits are well below the measured
extension of the Crab nebula. We note that Markarian$~$421 culminates
at large zenith angles of $\theta > 60^{\circ}$ at the \hess\ site,
making this source a particularly convincing test of our PSF
understanding: even under such challenging observation conditions,
Markarian$~$421 appears to be point-like. As we also show in
Fig.~\ref{fig_uls_ext}, we tested the Crab nebula dataset for a zenith
angle dependence by splitting the observations in two datasets above
and below $46^{\circ}$. The measured extensions are compatible with
each other.

Also shown in Fig.~\ref{fig_uls_ext} is the extension of the Crab
nebula cross-checked with an independent calibration, reconstruction,
and analysis method~\cite{2014_ImPACT}. We find this second extension
measurement slightly larger than our nominal value, and incorporate the
difference as one of our systematic uncertainties mentioned above.

\begin{figure*}
  \centering
  \includegraphics[width=0.49\textwidth]{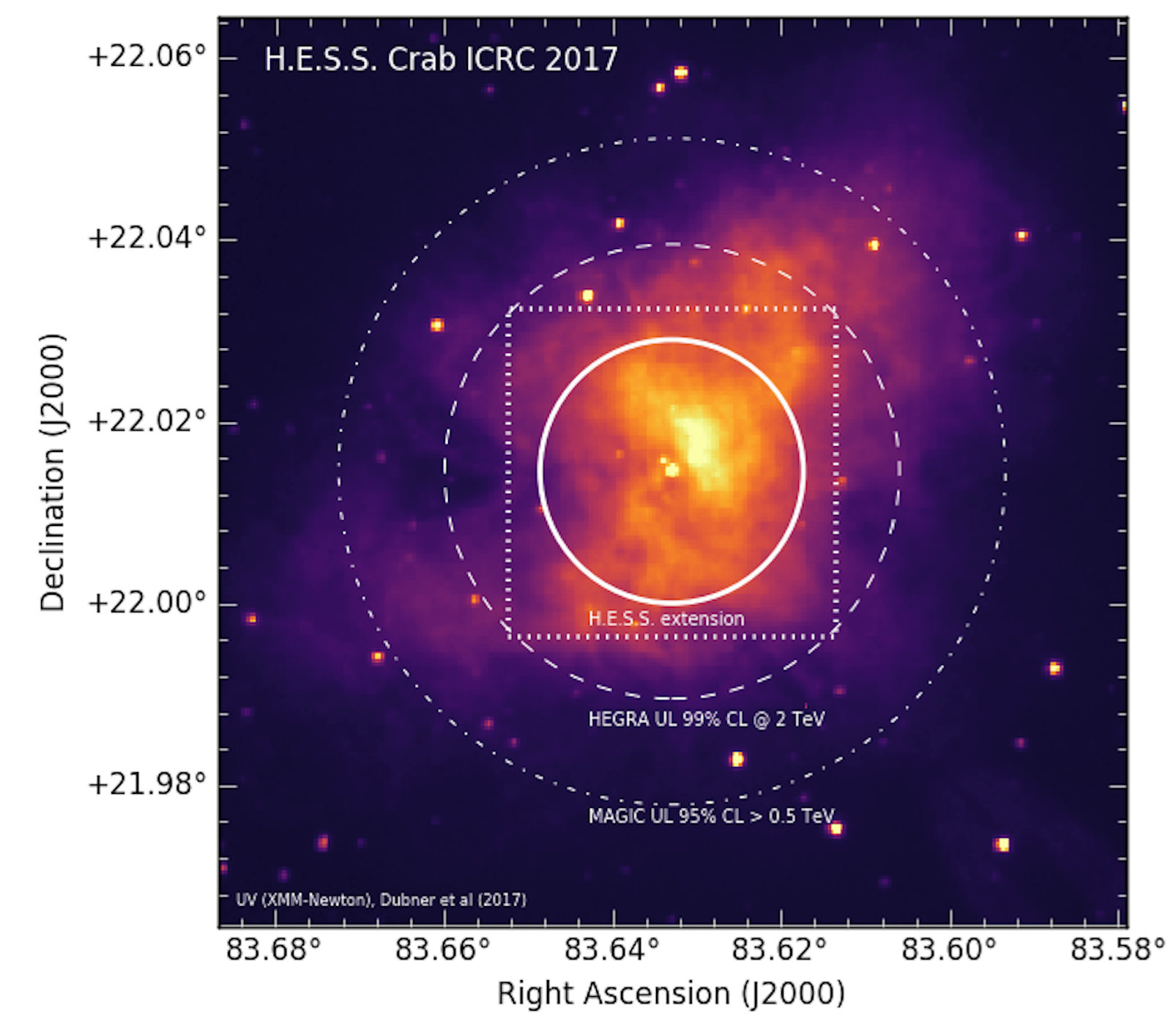}
  \includegraphics[width=0.475\textwidth]{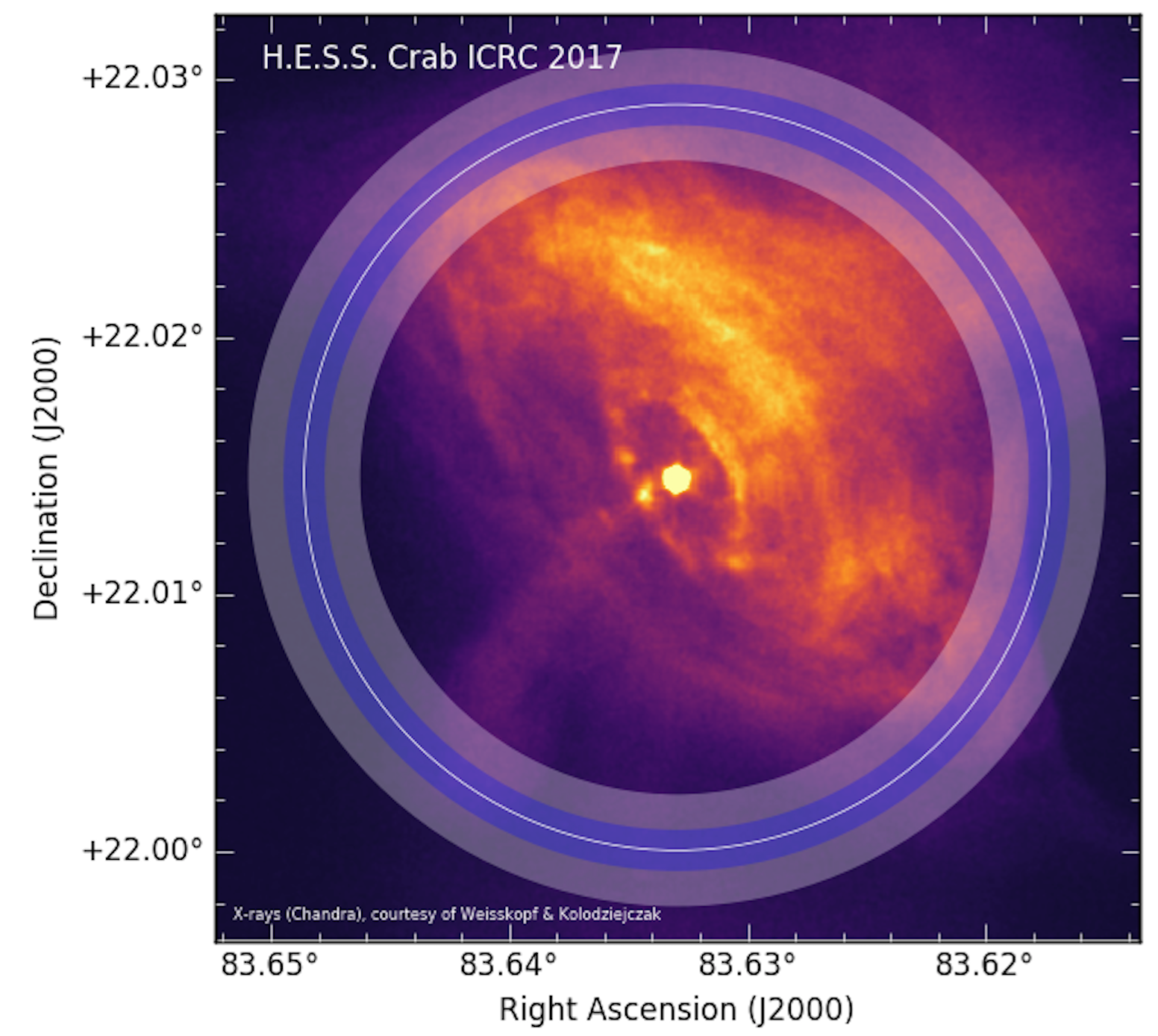}
  \caption{\textit{Left:} UV ($\lambda = 291\,$nm) image of the Crab
    nebula~\cite{2017_Dubner}. The MAGIC and HEGRA extension upper
    limits of $2.2'$~\cite{2008_MAGIC_Crab} and
    $1.5'$~\cite{2000_HEGRA_Crab} are drawn as dash-dotted and dashed
    lines, respectively. The extent of the sky region of the
    \textit{right} panel is indicated as dotted square, and the \hess\
    extension (Gaussian $\sigma$ as given in the main text) is drawn
    as a solid circle. All circles are centred on the Crab pulsar
    position. \textit{Right:} \emph{Chandra} X-ray image of the Crab
    nebula. The \hess\ extension is shown as solid white circle
    overlaid on top of shaded annuli indicating the statistical and
    systematic uncertainties of our measurement.}
  \label{fig_mwl}
\end{figure*}
Our VHE gamma-ray extension of the Crab nebula is compared to the
morphology found at UV wavelengths and X-ray energies in the
\textit{left} and \textit{right} panel of Fig.~\ref{fig_mwl},
respectively. The \hess\ extension covers a good fraction of the
optical nebula. Comparing the TeV gamma-ray emission from Inverse
Compton scattering to the keV synchrotron X-ray emission, we find that
the nebula measured with \hess\ is significantly larger than when
measured with \emph{Chandra}. This result is naturally explained by the
radiation cooling of electrons. The Crab nebula size therefore
decreases with increasing electron energy, and the energies of
electrons producing the TeV gamma rays are well below those of
electrons emitting the hard X-rays measured by \emph{Chandra}. The
measured size of the TeV gamma-ray nebula can be reproduced within the
standard magnetohydrodynamic model of Kennel and
Coroniti~\cite{1984ApJ...283..694K,1984ApJ...283..710K} assuming a 
magnetisation parameter $\sigma \approx 0.01$.

\section{Conclusions}
Here we document the ability of the \hess\ IACT array to robustly
measure VHE gamma-ray source extensions down to $30-40''$. The
performance boost provided by using time-dependent simulations allows
us to resolve for the first time the Inverse Compton component of the
Crab nebula. The emission region size we find is well below the
previously most constraining upper limit of
$1.5'$~\cite{2000_HEGRA_Crab} and is determined with a high accuracy
of $\approx 15\%$. Compared to the synchrotron Crab nebula seen in keV
X-rays, the VHE gamma-ray emission region is clearly more extended.

\section*{Acknowledgements}

We gratefully acknowledge the support from the agencies and institutions listed \href{https://www.mpi-hd.mpg.de/hfm/HESS/pages/publications/auxiliary/HESS-Acknowledgements-ICRC2017.html}{here}.

\bibliographystyle{JHEP}
\bibliography{sneak_preview}

\end{document}